\documentclass[twocolumn,showpacs,preprintnumbers,amsmath,amssymb]{revtex4}

\usepackage{graphicx}% Include figure files
\usepackage{dcolumn}% Align table columns on decimal point
\usepackage{amsfonts,amsmath,amssymb,bm}
\begin{document}

\title{Electric field response of strongly correlated one-dimensional metals: a Bethe-Ansatz density functional theory study}

\author{A. Akande and S. Sanvito \thanks{akandea@tcd.ie}}
\affiliation{School of Physics and CRANN, Trinity College, Dublin 2, Ireland}

\date{\today}

\begin{abstract}
We present a theoretical study on the response properties to an external electric field of strongly correlated one-dimensional metals. 
Our investigation is based on the recently developed Bethe-Ansatz local density approximation (BALDA) to the density functional 
theory formulation of the Hubbard model. This is capable of describing both Luttinger liquid and Mott-insulator correlations. 
The BALDA calculated values for the static linear polarizability are compared with those obtained by numerically accurate methods, 
such as exact (Lanczos) diagonalization and the density matrix renormalization group, over a broad range of parameters. In general 
BALDA linear polarizabilities are in good agreement with the exact results. The response of the exact exchange and correlation potential 
is found to point in the same direction of the perturbing potential. This is well reproduced by the BALDA approach, although the fine 
details depend on the specific parameterization for the local approximation. Finally we provide a numerical proof for the non-locality 
of the exact exchange and correlation functional. 
\end{abstract}

% insert suggested PACS numbers in braces on next line
% insert suggested keywords - APS authors don't need to do this

%\maketitle must follow title, authors, abstract, \pacs, and \keywords
\maketitle

\section{Introduction}
Material systems, whose electronic structure cannot be described at a mean field level, are conventionally named strongly correlated. 
These display an enormous variety of properties, which all originate from the interplay between Coulomb repulsion and kinetic energy, 
and from their dimensionality. Phenomena related to electron-electron correlation include metal-insulator transition, Tomonaga-Luttinger 
liquid behaviour and superconductivity, just to name a few \cite{Fazeka,Giamarchi}. In particular electron correlations play a fundamental 
r\^ole in one-dimension (1D). In 1D confined structures, electrons cannot avoid each other and collective excitations emerge over the 
ground state, so that the Fermi liquid picture breaks down. In fact one can demonstrate that the ground state of an interacting 1D object 
is always a Luttinger liquid regardless of the strength of the electron-electron interaction \cite{Fazeka,Giamarchi}. Although some aspects 
are still controversial, experimental evidence supporting the existence of Luttinger liquids in 1D has been provided for carbon 
nanotubes \cite{Postma} and for atomic wires built of surface terraces \cite{Segovia,Auslanender}. 

Strongly correlated systems are regularly modeled by means of effective Hamiltonians, which usually lack all the details of an 
{\it ab initio} description, but capture the relevant physical properties arising from electron correlation. The advantage of dealing 
with effective Hamiltonians is that they are commonly mathematically tractable and general enough to be applied to a variety of problems. 
Among the many effective Hamiltonians that one can construct the Hubbard model \cite{Gutzwiller, Hubbard, Kanamory} has enjoyed a 
vast popularity since it is simple and still can capture the subtle interplay between Coulomb repulsion and kinetic energy.

Although exact solutions of the Hubbard model are known in particular limits \cite{Hubbook}, a general one for an arbitrary system, which 
can be finite and inhomogeneous, requires a numerical treatment. This however represents a severely demanding task, since the Hilbert 
space associated to the Hubbard Hamiltonian for $L$ sites is 4$^L$ dimensional, so that exact (Lanczos) diagonalization (ED) can only handle 
a relatively small number of sites. Other many body approaches, such as the density matrix renormalization group (DMRG) 
\cite{White, Schollwock}, extend the range to a few hundred sites, but little is possible beyond that limit. It would be then useful to
have a method capable of describing accurately the ground state and still having the computational overheads of a mean field
approach. Such a method is provided by lattice density functional theory (LDFT). 

LDFT was initially proposed by Gunnarsson and Schonhammer \cite{Gunnarsson, Schonhammer} as an extension of standard, 
{\it ab initio}, DFT \cite{HK,KS} to lattice models. The theory essentially reformulates the Hohenberg-Kohn theorem and the Kohn-Sham 
construction in terms of the site occupation instead of the electron density. Although originally introduced with a pedagogical purpose, 
LDFT has enjoyed a growing success and it has been already applied to a diverse range of problems. These include fundamental 
aspects of DFT and of the Hubbard model, as the band-gap problem in semiconductors \cite{Gunnarsson}, the dimerization of 1D Hubbard 
chains \cite{Lopez} and the formation of the Mott-Hubbard gap \cite{Capelle1}. LDFT has also been employed for investigating effects 
at the nanoscale traceable to strong correlation, like the behavior of impurities \cite{Capelle2}, spin-density waves \cite{Capelle3} and 
inhomogeneity \cite{Silva}, as well as more exotic aspects like the phase diagram of harmonically confined 1D fermions \cite{Campo} 
and that of ultracold fermions trapped in optical lattices \cite{Xianlong1, Xianlong2, Xianlong3}. More recently LDFT has been 
extended to the time-dependent domain \cite{Verdozzi}, to quantum transport \cite{Gross} and to response theory \cite{Schenk}.

As in standard DFT also LDFT is in principle exact. However its practical implementation is limited by the accuracy of the unknown exchange 
correlation (XC) functional, which introduces the many-body effects into the theory. The construction of an XC functional begins with choosing 
a reference system, for which some exact results are known. These impose a number of constraints that the XC functional must satisfy,
as for example its asymptotic behavior or its scaling properties. Then the functional is built by interpolating and fitting to known
many-body reference results. Such a construction for instance has been employed in the case of the local density approximation 
(LDA) in {\it ab initio} DFT. The reference system in two and three dimensions is usually an electron gas of some kind, since one aims
at reproducing a Fermi liquid. However in 1D the known ground state has a Luttinger-liquid nature, so that the reference system should be
chosen accordingly. In the case of the Hubbard Hamiltonian in 1D a powerful result is that obtained by Lieb and Wu \cite{LiebWu} for the
homogeneous case by using the Bethe Ansatz. This is the basis for constructing an XC functional for the Hubbard model in
1D \cite{Capelle1,Capelle2}.

In this work we evaluate the ability of a range of known approximations to the XC functional for the 1D Hubbard model at predicting the electrical 
response to an external electric field of finite 1D chains away and in the vicinity of the Mott transition. This is relevant not just as a test for Hubbard
LDFT but also for understanding real materials, whose electrical response can be mimicked in terms of the Hubbard model 
\cite {Ishihara1, Ishihara2}. The 1D case in particular can provide important insights into the nonlinear optical properties of polymers \cite{Rojo}. 
Our strategy is that of constantly comparing the DFT results with those obtained with highly accurate many-body schemes. For these we
use exact diagonalization for small chains and the DMRG method for larger systems. Our calculations reveal a substantial good agreement 
between LDFT and exact results for both the polarizability and the XC potential response of finite 1D chains. The paper is organized as 
follows. In the next section we will briefly review the Hubbard LDFT and the approximations used for constructing the XC functional. Then 
we will discuss results, first for the electrical polarizabilities and then for the response of the XC potential to an external electric field. Finally we 
will carry on a numerical investigation on the validity of the local approximation to the XC functional and then we will conclude.

%%%%%%%%%%%%%%%%%%%%%%%%%%%%%%%%%%%%%%%%%%%%%%%%%%
%%%%%%%%%%%%%%%%%%%%%%%%%%%%%%%%%%%%%%%%%%%%%%%%%%

\section{Theoretical formulation}
\label{Theory}

One-dimensional correlated metals can be described by the homogeneous Hubbard Hamiltonian, $H_\mathrm{U}$. For a 1D chain comprising 
$L$ sites $H_\mathrm{U}$ writes
\begin{equation}\label{hub}
H_\mathrm{U}=-t \sum_{i=1,\:\sigma}^{L-1}(c^\dagger_{i+1\sigma}c_{i\sigma}+hc) + U\sum_{i=1}^L \hat{n}_{i\uparrow} \hat{n}_{i\downarrow}, 
\end{equation}
where the first kinetic term describes the hopping of electrons with spin $\sigma$ ($\sigma=\uparrow, \downarrow$) between nearest neighbour sites 
with amplitude $t>0$, while the second accounts for the electrostatic repulsion $U>0$ of doubly occupied sites. In the equation (\ref{hub}) 
$c_{i\sigma}^\dagger (c_{i\sigma})$ is the fermion creation (annihilation) operator for an electron at site $i$ with spin $\sigma$ and the site
occupation operator is written as $\hat{n}_{i\sigma}=c^\dagger_{i\sigma}c_{i\sigma}$. Clearly there is only one energy scale in the problem so
that the ratio $U/t$ determines all the electronic properties. Note that a second energy scale can be included in the problem by adding to the
Hamiltonian an on-site energy term $\sum_{i=1,\:\sigma}^L \epsilon_i\hat{n}_{i\sigma}$ mimicking a ionic lattice. 

As discussed in the introduction the fundamental quantity of LDFT is the site occupation, $n_i$, which is calculated by solving the equivalent Kohn-Sham
problem. This can be generally written as
\begin{equation}
 \sum_{j=1}^L [-t(\delta_{i+1\:j}+\delta_{i-1\:j})+v_\mathrm{KS}^{i}]\phi_j^{(\alpha)}=\epsilon^{(\alpha)}\phi_i^{(\alpha)}\;,
\label{KSeq}
\end{equation}
where $v_\mathrm{KS}^{i}$ is the general Kohn-Sham potential. The occupied Kohn-Sham eigenvectors, $\phi_i^{(\alpha)}$, define $n_i$
\begin{equation}
 n_i=\sum_{\alpha} w^{(\alpha)}|\phi_i^{(\alpha)}|^2\:,
\end{equation}
where $w^{(\alpha)}$ are the occupation numbers, which satisfy $\sum_\alpha w^{(\alpha)}=N$ with $N$ being the total number of electrons. By following 
in the footsteps of standard {\it ab initio} DFT the Kohn-Sham potential can be written as the sum of three terms
\begin{equation}\label{Pot}
v_\mathrm{KS}^{i}=[v_\mathrm{H}^i+v_\mathrm{ext}^i+v_\mathrm{XC}^i]\:,
\end{equation}
where $v_\mathrm{H}=Un_i/2$ is the Hartree potential and $v_\mathrm{ext}^i$ is the external one. The last term in equation (\ref{Pot}) is the
XC potential, which needs to be approximated. 

The Kohn-Sham equations simply follow by variational principle from the minimization of the energy functional. Thus the total energy  of the 
system, $E$, can be defined as
\begin{equation}\label{EnFunc}
E[\{n_i\}]=\sum_\alpha w^{(\alpha)}\epsilon^{(\alpha)}-\sum_iv_\mathrm{XC}^in_i-\sum_i\frac{Un_i^2}{4}+E_\mathrm{XC}[\{n_i\}]\:,
\end{equation}
where the last term is the XC energy. Note that different values of $U$ and $t$ define completely the theory, so that one has a different
functional for every value of $U/t$.

We now review the strategy used for constructing a suitable local $v_\mathrm{XC}^i$ \cite{Capelle1,Capelle2,Capelle3}. The guiding idea 
is that of defining $v_\mathrm{XC}^i$ as the local counterpart of the Bethe Ansatz potential for the homogeneous Hubbard model (for an
infinite number of sites), i.e.
\begin{equation}\label{BALDA}
 v_\mathrm{XC}^i|_\mathrm{BALDA}=v_\mathrm{XC}^\mathrm{hom}(n,t,U)|_{n\rightarrow n_i}\:.
\end{equation}
Here BALDA stands for Bethe Ansatz local density approximation and $v_\mathrm{XC}^\mathrm{hom}(n,t,U)$ is the XC potential for 
the homogeneous Hubbard model, which is defined only in terms of the band filling $n=N/L$ (note that for the homogeneous case 
$n_i=n$ for every site $i$). Formally, and in complete analogy with {\it ab initio} DFT, $v_\mathrm{XC}^\mathrm{hom}(n,t,U)$ is obtained 
by functional derivative of the exact energy density, $e(n,t,U)$, of the reference system (in this case the homogeneous Hubbard 
model), after having subtracted the kinetic energy density of the non-interacting case $e(n,t,U=0)$ and the Hartree energy density, 
$e_\mathrm{H}(n,U)$. This gives us
\begin{equation}
v_\mathrm{XC}^\mathrm{hom}(n,t,U)=\frac{\partial}{\partial n}[e(n,t,U)-e(n,t,U=0)-e_\mathrm{H}(n,U)]\;.
\end{equation}

The question is now how to obtain $e(n,t,U)$. Two alternative constructions have been proposed in the past and here we have 
adopted and numerically implemented both. The first one consists in using the analytical parameterization proposed by Lima 
et al. \cite{Capelle2, Capelle3}, which interpolates the known exact results for: 1) $U\rightarrow 0$ and any $n\le1$, 2) 
$U\rightarrow\infty$ and any $n\le1$ and 3) $n=1$ and any $U$. The resulting XC potential can then be written as 
\begin{equation}
v_\mathrm{XC}^\mathrm{hom}(n,t,U)=t\mu\left[2 \cos\frac{k\pi}{\beta(U)}-2 \cos\frac{k\pi}{2}+\frac{kU}{2}\right],
\end{equation}
where $k=1-|n-1|$, $\mu=$ sgn$(n-1)$ and $\beta(U)$ is a $U$-dependent parameter, which can be determined by solving a
transcendental equation. The alternative route is that of employing a direct numerical solution of the coupled Bethe Ansatz integral equations. 
This approach has been already used for the study of ultracold repulsive fermions in 1D optical lattices \cite{Xianlong1}. 

\begin{figure}[htb]
\centering{\includegraphics[width=0.92\linewidth]{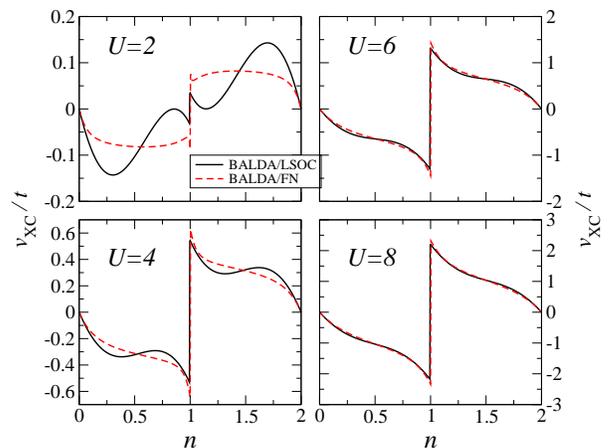}}
\caption{(Color online) Exchange-correlation potential $v_\mathrm{XC}^\mathrm{hom}(n,t,U)$ of the 1D Hubbard model as a function of the electron filling, $n$,
for different values of interaction strength, $U$. Here we report data for both BALDA/LSOC and BALDA/FN. Note that the agreement between the two
schemes improves as $U$ increases.} 
\label{xcpotential}
\end{figure}

The first parameterization is known as BALDA/LSOC \cite{LSOC} and the second as BALDA/FN (FN~=~fully numerical). In figure~\ref{xcpotential} 
the XC potentials as a function of the electron filling for the two schemes are shown for different values of $U$. In the picture (and in the calculations) 
we always use the particle-hole symmetry, which imposes $v_\mathrm{XC}^\mathrm{hom}(n>1,t,U)=-v_\mathrm{XC}^\mathrm{hom}(2-n,t,U)$. 
From the figure one can immediately observe that the potential in both cases has a discontinuity in the derivative at half-filling ($n=1$, $N=L$).
This reflects the fact that the underlying homogeneous 1D Hubbard model has a metal-insulation transition for $n=1$. Such a discontinuity
in the derivative of the potential, as in standard {\it ab initio} DFT, is responsible for the opening of the energy gap. The second observation is that
the two parameterizations always coincide by construction at $n=0$ and $n=2$ but that their agreement over the entire $n$ range depends
on the value of $U$. In particular one can report a progressively good agreement as $U$ increases. This is not a surprise since the BALDA/LSOC
potential is constructed to exactly reproduce the $U\rightarrow\infty$ limit.
%

%%%%%%%%%%%%%%%%%%%%%%%%%%%%%%%%%%%%%%%%%%%%%%%%%%%%%%%%%%%%%%%%%%%%%%%%%%%%%%%%%%%%%
%%%%%%%%%%%%%%%%%%%%%%%%%%%%%%%%%%%%%%%%%%%%%%%%%%%%%%%%%%%%%%%%%%%%%%%%%%%%%%%%%%%%

\section{Polarizabilities}
\label{Polarizabilities}

We calculate the electrical polarizability of linear chains with the finite difference method, i.e. as numerical derivative of calculations performed 
at different external electric fields. An external electric field enters into the problem by adding to the Hubbard Hamiltonian $H_\mathrm{U}$ the term
\begin{equation}
H_{\cal E}=e{\cal E}\hat{x}=e{\cal E}\sum_{i=1}^L(i-\bar{x})c^\dagger_{i}c_{i}\:,
\label{hamilEF}
\end{equation}
where $\bar{x}=\frac{1}{2}(L+1)$ is the middle site position of the chain, $e$ is the electronic charge ($e=-1$) and ${\cal E}$ is the electric
field intensity (the electric field is applied along the chain). 
In general the electrical dipole, $P$, induced by an external electric field can be calculated simply as the expectation value of the dipole
operator over the ground state wave-function $|\Psi_0({\cal E})\rangle$ (note that this is a general definition so that $|\Psi_0({\cal E})\rangle$ 
is not necessarily the Kohn-Sham ground-state wave-function), i.e.
\begin{equation}
P=e\langle \Phi_0({\cal E})|\sum_{i=1}^L(i-\bar{x})c^\dagger_{i}c_{i}|\Phi_0({\cal E})\rangle=\frac{\mathrm{d}E_0({\cal E})}{\mathrm{d}{\cal E}},
\end{equation}
where $E_0$ is the ground state energy. For small fields $P$ can be Taylor expanded about ${\cal E}=0$ so that the linear polarizability, 
$\alpha$, is defined as
\begin{equation}
P\sim\alpha{\cal E}+\gamma{\cal E}^3+ {\cal O}({\cal E}^5)\:,\;\;\;\;\;\;\;\;\alpha=\frac{\mathrm{d^2}E_0({\cal E})}{\mathrm{d}{\cal E}^2}\:.
\label{polar}
\end{equation}
Our calculation then simply proceeds with evaluating $E_0({\cal E})$ for different values of ${\cal E}$ and then by fitting the first derivatives 
with respect to the field to the equation (\ref{polar}), as indicated in reference \cite{Rojo}. We note that our finite difference scheme is not 
accurate enough for calculating the hyper-polarizability, $\gamma$, which then is not investigated here. 

It has been already extensively reported that BALDA-LDFT gives a substantial good agreement with exact calculations in terms of ground 
state total energy \cite{Capelle1,Capelle2}. The polarizability however offers a more stringent test for the theory since it involves derivative of $E_0$.
Hence it is important to compare the various approximations with exact results.
\begin{figure}[htb]
\centering{\includegraphics[width=1.0\linewidth]{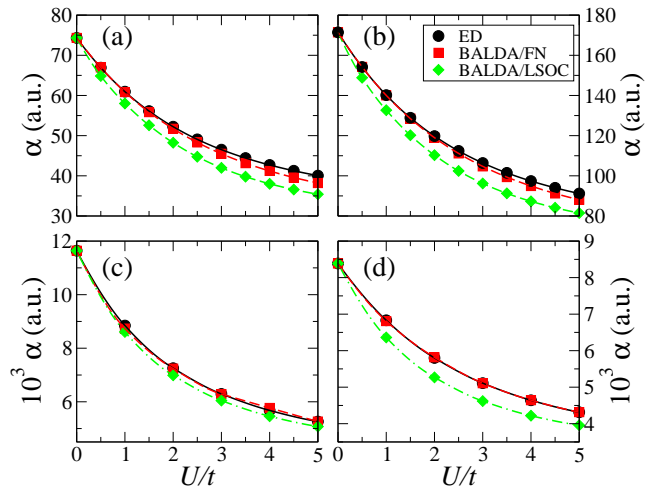}}
\caption{(Color online) Linear polarizability, $\alpha$, as a function of the Coulomb repulsion $U/t$. Results are presented for BALDA/LSOC and 
BALDA/FN and they are compared with those obtained with either exact diagonalization (ED) or DMRG calculations. In the various panels we show: 
(a) $L=12$ at quarter filling ($n=1/2$), (b) $L=16$ at quarter filling, (c) $L=60$ and $N=20$, and (d) $L=60$ at quarter filling.}\label{jointpolar}
\end{figure}
For small chains, $L<18$, these are obtained by simply performing ED. However for the longer chains
ED is no longer feasible and we employ instead the DMRG scheme \cite{Schollwock, DMRG}. DMRG has been 
widely used to investigate one-dimensional and quasi one-dimensional quantum systems. It usually performs best with open boundary 
conditions and utilizes appreciable computational resources depending on the number of states that are kept for the calculation. Our DMRG 
calculations are performed by employing the Algorithms and Libraries for Physics Simulations (ALPS) \cite{ALPS} package for strongly 
correlated quantum mechanical systems. The DMRG results are obtained by using a cutoff of $m=350$, i.e. by retaining the dominant 350 
density matrix eigenvectors.

Let us start our analysis by looking at the polarizability as a function of the energy scale $U/t$. Selected results for quarter-filling, $n=1/2$, 
and for $n=1/3$ are presented in the various panels of figure \ref{jointpolar}. Note that throughout this work we always stay away from the 
half-filling case ($n=1$), where the derivative discontinuity of the potential makes the LDFT convergence problematic.

In general we find that the polarizability decreases monotonically with increasing the on-site repulsion $U$. This is indeed an expected result
since an increase in on-site repulsion means a suppression of charge fluctuations and consequently a reduction of $\alpha$. Away from $U=0$
the dependence of $\alpha$ on $U/t$ can be fitted with
\begin{equation}
\alpha(U/t;L,n)=\alpha_0(L,n)\left(\frac{U}{t}\right)^{-\xi(L,n)}\:,
\label{alphaU}
\end{equation}
where all the parameters have a dependance on the length of the chain and on the band filling. The results of such a fitting procedure are 
reported in table~\ref{Tab1}. Note that in the fit we did not impose any constraints and we have included only points with $U/t\ge1$.
 \begin{table}[h]
\begin{tabular}{cccccc} \hline\hline
Method & $L$ & $N$ & $n$ & $\alpha_0$ & $\xi$ \\ \hline
{\sc ed} & 12 & 6 & 1/2 & 59.69 & 0.23 \\ 
{\sc balda/lsoc} & & & & 62.06 & 0.27 \\ 
{\sc balda/fn}     & & & & 59.50 &  0.25 \\ \hline
{\sc ed} & 16 & 8 & 1/2 & 142.88 & 0.27 \\ 
{\sc balda/lsoc}   & & & & 135.07 & 0.31 \\ 
{\sc balda/fn}       & & & & 143.56 & 0.30 \\ \hline
{\sc dmrg} & 60 & 30 & 1/2 & 8939.5 & 0.32 \\ 
{\sc balda/lsoc}          & & & & 8673.1 & 0.33 \\ 
{\sc balda/fn}              & & & & 8837.8 & 0.31 \\ \hline
{\sc dmrg} & 60 & 20 & 1/3 & 6931.6 & 0.29 \\ 
{\sc balda/lsoc}         & & & & 6401.0 & 0.30 \\ 
{\sc balda/fn}             & & & & 6920.7 & 0.29 \\ \hline
\hline\hline
\end{tabular}\caption{\label{Tab1}Scaling parameters for $\alpha(U/t;L)$ as obtained by fitting the data of Fig.~\ref{jointpolar} to the expression
of equation (\ref{alphaU}). Note that the fit has been obtained without any constraints and by including data only for $U/t\ge1$. }
\end{table}

\begin{figure}[htb]
\centering{\includegraphics[width=1.0\linewidth]{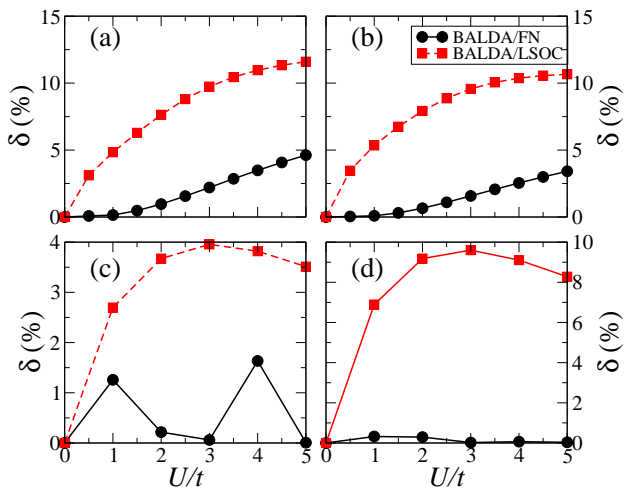}}
\caption{(Color online) Relative error between BALDA calculated polarizabilities and those obtained with exact methods (either ED or DMRG).
In the panels we show: (a) $L=12$ at quarter filling ($n=1/2$), (b) $L=16$ at quarter filling, (c) $L=60$ and $N=20$, and (d) $L=60$ 
at quarter filling.} \label{jointpolarerror}
\end{figure}

From the fit and from figure~\ref{jointpolar} one can immediately note that both the BALDA flavors of the exchange and correlation functional
reproduce rather well the exact results, in good agreement with previously published calculations \cite{Schenk}. The agreement is particularly 
good for the FN functional, which matches the ED/DMRG results almost perfectly over the entire range of $U/t$'s and filling investigated. A 
quantitative assessment of goodness of the BALDA results is provided in figure~\ref{jointpolarerror} where the relative error, $\delta$, from 
the reference exact calculations is presented. In general, and as expected, we find that the error grows with $U/t$, i.e. with the system 
departing from the non-interacting case. However, there is also a saturation of the error as the interaction strength increases, reflecting the 
fact that both the BALDA potential are exact in the limit of $U\rightarrow\infty$. As a further consequence of the $U\rightarrow\infty$ limit, we 
also observe that the relative error between BALDA/LSOC and BALDA/FN reduces as $U$ grows. 

Given the accuracy of the BALDA/FN scheme we have decided to use the same to investigate in more details the scaling properties of 
$\alpha(U/t;L)$. First we look at the scaling as a function of the interaction strength $U/t$. In this case we always consider a chain 
containing $L=60$ sites for which the deviation from the DMRG results is never larger than 2\%. Furthermore this is a length which 
allows us to explore a rather large range of electron filling, so that it allows us to gain a complete understanding of the scaling properties.  
\begin{figure}[htb]
\centering{\includegraphics[width=1.0\linewidth]{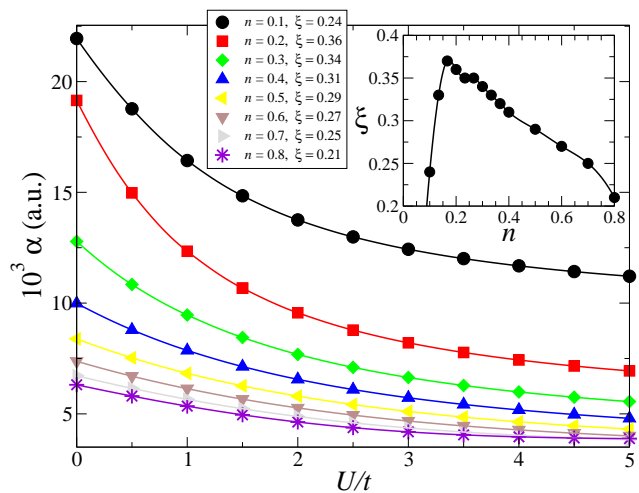}}
\caption{(Color online) Polarizability as a function of $U/t$ for a chain of 60 sites and various filling factors, $n$. The figure legend reports
the fitted values for the exponent $\xi$ [see equation (\ref{alphaU})]. The symbols represents the calculated data while the solid lines are just 
to guide the eyes. In the inset we present the exponent $\xi$ as a function of the filling factor $n$.} 
\label{scalingU}
\end{figure}
Our results are presented in figure~\ref{scalingU} where we show $\alpha$ as a function of $U/t$ for different filling factors, we list the values 
of $\xi$ obtained by fitting the actual data for $U/t\ge1$ to the expression in equation (\ref{alphaU}) and we provide (inset) the dependence
of $\xi$ on $n$.

In general the fit to our data is excellent, suggesting the validity of the exponential scaling of the polarizability with the interaction strength 
(away from half filling). In particular we find that $\xi$ decreases monotonically with $n$ for $n>0.2$ but it increases for smaller values. 
This means that $\xi(n)$ has a maximum just before $n=0.2$, which appears rather sharp (see inset of figure~\ref{scalingU}). We are at
present uncertain about the precise origin of such a non-monotonic behavior. However, as we will see in details later on, we notice that 
the response of the exchange and correlation potential to the external electric field has an anomaly for small $U$ and $n$. We believe 
that such an anomaly might be the cause of the non-monotonic behaviour of $\xi$.

\begin{figure}[htb]
\centering{\includegraphics[width=1.0\linewidth]{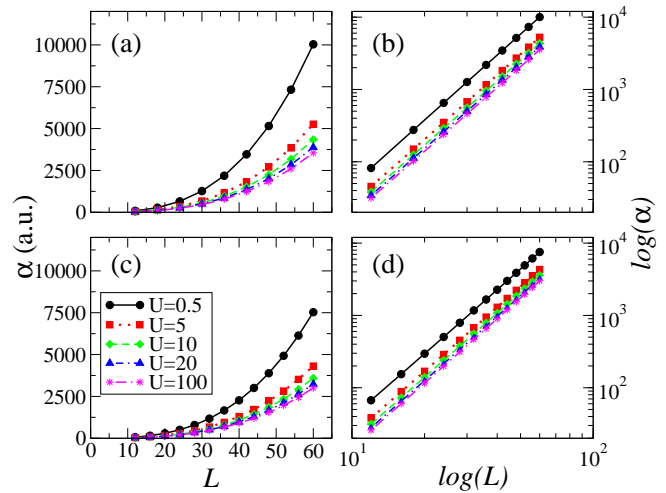}}
\caption{(Color online) Scaling of the polarizability as a function of the chain length, $L$. Panel (a) and (b) are for $n=1/3$ while
(c) and (d) for $n=1/2$. Note the linear dependence of the $\alpha(L)$ curve when plotted on a log-log scale, proving the relation
$\alpha(L)=\alpha_1 L^{\gamma}$} 
\label{Fig5}
\end{figure}

Next we turn our attention to the scaling of $\alpha$ with the chain length. In figure \ref{Fig5} we present $\alpha(L)$ for two different 
filling factors ($n=1/3$ and $1/2$) and different values of $U/t$. 
Data are plotted both in linear and logarithmic scale, from which a clear power-law dependance of $\alpha$ on $L$ emerges. A
fit to our data provides the following scaling
\begin{equation}
\alpha(U/t;L)=\alpha_1L^{\gamma}\:.
\label{alphaL}
\end{equation}
Importantly this time we find essentially no dependance of both $\alpha_1$ and $\gamma$ on either $U/t$ or $n$. The fit reveals a value for the
exponent of $\gamma\sim3$ (the range is from $\gamma=2.93$ to $\gamma=2.98$). This is what expected for free electrons in 1D \cite{Rojo}, and
it is substantially different from the predicted linear scaling at $n=1$. Our results thus confirms that away from $n=1$ the electrostatic response
of the Hubbard model is similar to that of the non-interacting electron gas. Going in more details we find a rather small monotonic dependance 
of $\gamma$ on $U/t$. This however depends also on $n$ since for $n=1/3$ we find that $\gamma$ reduces as $U/t$ is increased (from 2.98 
for $U/t=0.5$ to 2.93 for $U/t=100$), while the opposite behavior is found for $n=1/2$ ($\gamma=2.94$ for $U/t=0.5$ and 2.96 for $U/t=100$).

%%%%%%%%%%%%%%%%%%%%%%%%%%%%%%%%%%%%%%%%%%%%%%%%%%
%Exchange potential wrt to field
%%%%%%%%%%%%%%%%%%%%%%%%%%%%%%%%%%%%%%
\section{Response of the BALDA potential to the external field}

In {\it ab initio} DFT the failures of local and semi-local XC functionals in reproducing accurate linear polarizabilities are related to the 
incorrect response of the XC potential to the external electric field \cite{Gisbergen, Perdew}, which in turn originates from the presence of 
the self-interaction error \cite{Kummel,Das}. In particular for {\it ab initio} DFT the exact XC potential should be opposite to the external 
one, while the LDA/GGA (generalized gradient approximation, GGA) returns a potential which responds in the same direction. In order to 
investigate the same feature for the case of the Hubbard model LDFT we calculate the potential response 
\begin{equation}
\Delta v_\mathrm{XC}=v_\mathrm{XC}^{\cal E}(n_i)-v_\mathrm{XC}^{{\cal E}=0}(n_i)\:,
\end{equation}
where $v_\mathrm{XC}^{\cal E}(n_i)$ is the exchange and correlation potential at site $i$ in the presence of an electric field ${\cal E}$.
Also in this case we adopt the finite difference method and we use ${\cal E}=0.01$, after having checked that the trends remaining 
unchanged irrespectively of the field strength. 

In order to provide a benchmark for our calculations we also need to evaluate the potential response for the exact Hubbard model. 
We construct the exact potential by reverse engineering, a strategy introduced first by Almbladh and Pedroza \cite{Almbladh} and by 
von~Barth \cite{vbar} and then applied to both static and time dependent LDFT by Verdozzi \cite{Verdozzi}. This consists in minimizing 
about the Kohn-Sham potential the functional ${\cal F}$ (in reality here this is just a function) defined as
\begin{equation}
 {\cal F}[v_\mathrm{XC}]=\sum_i^L(n_i^\mathrm{KS}-n_i^\mathrm{exact})^2,
\end{equation}
where $n_i^\mathrm{exact}$ is the exact site occupation at site $i$ as obtained by either ED or the DMRG method, while 
$n_i^\mathrm{KS}$ is the Kohn-Sham one. 

\begin{figure}[htb]
\centering{\includegraphics[width=0.48\textwidth,angle=0.0]{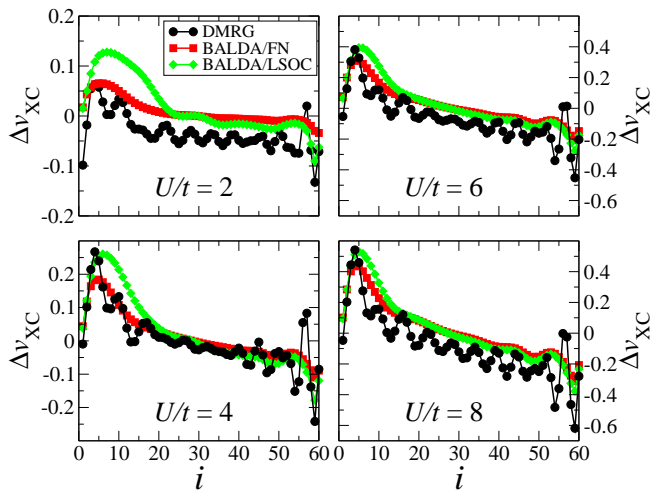}}
\caption{(Color online) The difference, $\Delta v_\mathrm{XC}$, between the XC potential calculated at finite electric field and in absence of the field
as a function of the site index. Results are presented for a 60 site chain with $N=10$ ($n=1/6$). The dots are the calculated data while the lines are 
a guide to the eye. The external potential has a negative slope.} 
\label{delta_exchange_60_10_LSOC_FN}
\end{figure}

\begin{figure}[htb]
\centering{\includegraphics[width=0.48\textwidth,angle=0.0]{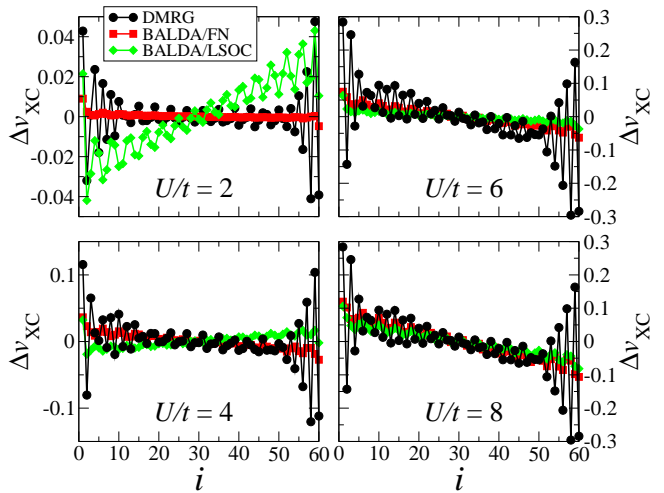}}
\caption{(Color online) The difference, $\Delta v_\mathrm{XC}$, between the XC potential calculated at finite electric field and in absence of the field
as a function of the site index. Results are presented for a 60 site chain with $N=30$ ($n=1/2$). The dots are the calculated data while the lines are 
a guide to the eye. The external potential has a negative slope.} 
\label{delta_exchange_60_30_LSOC_FN}
\end{figure}

Our results are summarized in figures~\ref{delta_exchange_60_10_LSOC_FN} and \ref{delta_exchange_60_30_LSOC_FN}, where we 
show $\Delta v_\mathrm{XC}$ as a function of the site index for a 60 site chain occupied respectively with 10 ($n=1/6$) and 30 ($n=1/2$)
electrons. The external electrostatic potential here decreases as the site number increases, i.e. it has a negative slope. Results are presented
for DMRG, BALDA/LSOC and BALDA/FN and for different values of $U/t$. 

In general and in contrast with {\it ab initio} DFT, we find that the response of the exact Hubbard-LDFT XC potential is in {\it the same direction} of 
the external perturbation for both the filling factors investigated and regardless of the magnitude of $U/t$. The response however becomes larger
as $U/t$ is increased (the slope of $\Delta v_\mathrm{XC}$ is more pronounced), a direct consequence of the fact that for large $U$'s small 
deviations from an homogeneous charge distribution produce large fluctuations in the potential. Such a behaviour is well reproduced by both the 
BALDA functionals, with the BALDA/FN scheme performing marginally better than the BALDA/LSOC one, and reflecting the same trend 
already observed for the polarizabilities. 

There is however one anomaly in the potential response for the BALDA/LSOC functional, namely at $n=1/2$ and for small $U/t$ (respectively 2 and 4)
the potential response is actually opposite (positive slope) to that of the DMRG benchmark. This means that in these particular range of filling and
interaction strength the BALDA/LSOC potential erroneously opposes to the external perturbation. The anomaly originates from the particular shape
of the BALDA/LSOC potential as a function of $n$ for small $U/t$ (see figure~\ref{xcpotential}). In fact, $v_\mathrm{XC}^i$ for BALDA/LSOC has a 
minimum for both $U/t=2$ and $U/t=4$ at around $n=1/4$, which means that its slope changes sign when the occupation sweeps across $n=1/4$.
Therefore for those critical interaction strengths the response is expected to be along the same direction of the external potential for $n<1/4$ and
for $3/4\lesssim n\le 1$ and opposite to it for $1/4<n\lesssim3/4$ (at $n\sim3/4$ there is a second change in slope). 

In the case of the BALDA/FN functional such an anomaly is in general not expected, except for small $U/t$ and $n$ close to the discontinuity at
$n=1$ (see figure~\ref{xcpotential}). This, however, is in the range of occupation not investigated here. Nevertheless we note that for $n=1/2$ 
and $U/t=2$ the BALDA/FN $v_\mathrm{XC}$ is almost flat. This feature is promptly mirrored in the potential response of figure~\ref{delta_exchange_60_30_LSOC_FN}, which also shows an almost flat $\Delta v_\mathrm{XC}$, although still with the correct negative slope. 

Given the good agreement for both the polarizability and the potential response between the exact results and those obtained with the BALDA 
(in particular with the FN flavour), one can conclude that the local approximation to the Hubbard-LDFT functional is adequate. Still it
is interesting to assess whether the remaining discrepancies have to do with the particular local parameterization of $E_\mathrm{XC}[\{n_i\}]$,
or with the fact that the exact XC functional may be intrinsically non-local. In order to answer to this question we have set a numerical test. We consider
a 60 site chain with $n=1/2$ (this should be long enough to resemble the infinite limit) and we introduce a local perturbation in half of the
chain. This is in the form of a reduction of the on-site energy of the first 30 sites by $\delta$. We then calculate the deviation of the XC potential 
$\delta v$ as a function of the deviation of the total energy $\delta E_0$. These two quantities are defined respectively as 
\begin{equation}
\delta v=\sum_i|v_\mathrm{XC}^{\delta}(n_i)-v_\mathrm{XC}^{\delta=0}(n_i)|\:,\;\;\;\;\;\;\;
\delta E_0=E_0(\delta)-E_0(0)\:,
\end{equation}
with $v_\mathrm{XC}^{\delta}$ and $E_0(\delta)$ respectively the XC potential at site $i$ and the total energy calculated for $\delta\ne0$. 
One then expects for a local potential that $\delta v\rightarrow 0$ as $\delta E_0\rightarrow 0$.

\begin{figure}[htb]
\centering{\includegraphics[width=0.44\textwidth,angle=0.0]{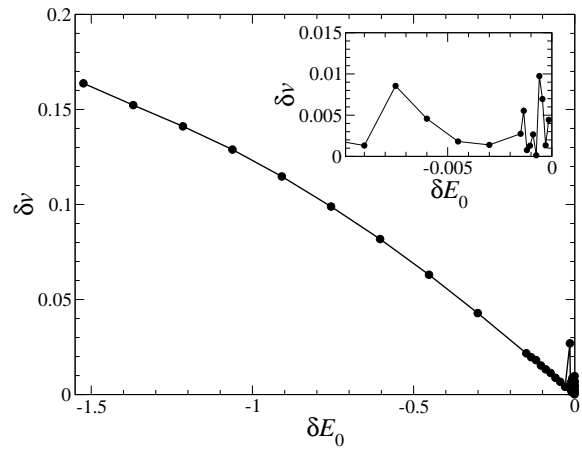}}
\caption{Variation of the XC potential, $\delta v$, as a function of the variation of the total energy, $\delta E_0$, for a 60 site chain in which 
the first 30 sites have an on-site energy lower by $\delta$ with respect to the remaining 30. The variation are calculated with respect to the 
homogeneous case. The inset shows a magnification of the data for small $\delta E_0$.} 
\label{energydifference}
\end{figure}

Our results are presented in figure \ref{energydifference}. These have been obtained for a relatively small $U/t=2$ by varying $\delta$ in the range
$0\le \delta \le 0.1$ in steps of 10$^{-5}$ (this range is used only for small $\delta$, while a coarse mesh is employed for large $\delta$). Interestingly 
we note that, after a steady decrease of $\delta v$ with reducing $\delta E_0$, the deviation of the potential starts to fluctuate independently on the 
size of $\delta E_0$. We have carefully checked that such fluctuations are well within our numerical accuracy, so that they should be attributed 
to the breakdown of the local approximation. We then conclude that part of the failure of BALDA/FN in describing the polarizability of finite 1D chains
must be ascribed to the violation of the local approximation.
%

%%%%%%%%%%%%%%%%%%%%%%%%%%%%%%%%%%%%%%%%%%
\section{Conclusions}
In conclusion, we have reported a systematic study of the electrical response properties of one-dimensional metals described by the 
Hubbard model. This is solved within LDFT and local approximations of the exchange and correlation functional. Whenever possible 
the calculations are compared with exact results obtained either by exact diagonalization of with the density matrix renormalization 
group approach. In general we find that BALDA functionals perform rather well in describing the electrical polarizability of finite 
one-dimensional chains. The agreement with exact results is particularly good in the case of numerically evaluated functionals. A similar 
good agreement is found for the exchange and correlation potential response. In this case we obtain the interesting result that the potential 
response is always along the same direction of the perturbing potential, in contrast to what happens in {\it ab initio} DFT. Furthermore for 
small electron filling and weak Coulombic interaction the commonly used LSOC parameterization is qualitatively incorrect due to a spurious 
minimum in the potential as a function of the site occupation. Finally we provide a numerical test of the breakdown of the local approximation 
being the source of the remaining errors.

\section{Acknowledgements}
A.A. thanks N. Baadji, I. Rungger and V. L. Campo for useful discussions. This work is supported by Science Foundation of Ireland under the grant SFI05/RFP/PHY0062 and 07/IN.1/I945. Computational resources 
have been provided by the HEA IITAC project managed by the Trinity Center for High Performance Computing and by ICHEC.


\begin{thebibliography}{99}
\bibitem{LSOC} LSOC is after the name of Lima, Silva, Oliviera and Capelle, who proposed the approximation.
\end{thebibliography}
\end{document}